\documentclass{acm_proc_article-sp}

\usepackage{epstopdf}
\usepackage{subfigure}
\usepackage{graphicx}
\usepackage{xcolor,colortbl}

\begin{document}

\title{Supporting Sensing Application in Vehicular Networks}

\numberofauthors{2} 
\author{
\alignauthor
Mohammad Nozari Zarmehri\titlenote{M. Nozari's work is supported by the Funda\c c\~ ao para Ci\^ encia e Tecnologia under the grant SFRH/BD/71438/2010 and by the CMU|Portugal project "DRIVE-IN: Distributed Routing and Infotainment through VEhicular Inter-Networking"}\\
       \affaddr{Instituto de Telecomunica\c c\~ es (IT),\\
	Faculdade de Engenharia da Universidade do Porto (FEUP)}\\
       \affaddr{Rua Dr. Roberto Frias, 4200-465}\\
       \affaddr{Porto Portugal}\\
       \email{mohammad.nozari@fe.up.pt}
\alignauthor
Ana Aguiar\\
       \affaddr{Instituto de Telecomunica\c c\~ es (IT),\\
	Faculdade de Engenharia da Universidade do Porto (FEUP)}\\
       \affaddr{Rua Dr. Roberto Frias, 4200-465}\\
       \affaddr{Porto Portugal}\\
       \email{ana.aguiar@fe.up.pt}
}

\maketitle
\begin{abstract}
This research aims at using vehicular ad-hoc networks as infra-structure for an urban cyber-physical system in order to gather data about a city. In this scenario, all nodes are data sources and there is a gateway as ultimate destination for all packets. Because of the volatility of the network connections and uncertainty of actual node placement, we argue that a broadcast-based protocol is the most adequate solution, despite the high overhead. 

The Urban Data Collector (UDC) protocol has been proposed which uses a distributed election of the forwarding node among the nodes receiving the packet: nodes that are nearer to the gateway have shorter timers and a higher forwarding probabilities. The performance of the UDC protocol has been evaluated with different suppression levels in terms of  the amount of collected data from each road segment using NS-3, and our results show that UDC can achieve significantly higher sensing accuracy than to other broadcast-based protocols.
\end{abstract}

\section{Introduction}
Vehicular ad-hoc networks (VANET) have moved into the spotlight driven mainly by the benefits expected from safety, traffic management and infotainment applications~\cite{Gerla2010}. In this paper we propose a different utilization, namely to use a VANET as the infra-structure for an urban monitoring system, an approach that has not been explored so far. Vehicles equipped with a wide range of sensing devices and the ability to communicate with each other offer a unique opportunity for gathering real-time data about a city, like traffic conditions~\cite{Hao2010}, environmental parameters, video and audio for surveillance~\cite{Gerla_2011}, or physical condition of the drivers~\cite{Rodrigues2010}. Knowing the updated state of relevant variables for a city is not only critical for applications such as navigation using real-time traffic information both for regular and for emergency vehicles, but also for personal mobility support and environmental monitoring~\cite{Rodrigues2011}.

The IEEE has recently released the 802.11p standard for VANET~\cite{IEEE2010}, which is derived from the 802.11a and 802.11e standards. The 802.11p PHY uses the 5.9~GHz frequency band with channels of 10~MHz. Additionally, 802.11p MAC does not require node association for communication, and no handshakes or acknowledgments are foreseen on the control channel.

The purpose of urban sensing is to provide information about an urban area to entities outside the VANET in near real-time. Each node collects information about its environment and sends it periodically to the gateway, configuring a many-to-one network topology, where all nodes are simultaneously both data generators and forwarders and the ultimate destination of the data lies outside the VANET.

Several protocols for data collection in wireless sensor networks (WSN) and mobile adhoc networks (MANETs) have been previously proposed, but VANET links are more volatile and multi-hop paths have very short durations~\cite{viriyasitavat_jsac2011}, in addition to the lack of reliability at the link level due to the highly dispersive environment. In this scenario, the protocols developed for WSN incur a very high path management overhead~\cite{Li2009} and are not adequate. On the other hand, existing VANET protocols focus on unicast or data dissemination, and the scenario described has not been previously addressed.

This article proposes and evaluates a data collection protocol over an urban VANET. 
Our main contributions are 1) a new data collection protocol for data gathering in urban area; 2) the evaluation of different suppression levels to limit the amount of forwarders at each hop; 3) the benchmarking of broadcast-based protocols using a metric appropriate to measure sensing performance.

Our results show that significant gains can be obtained from probabilistic forwarding to further reduce the replication of packets at each hop while there are significant amount of packets from each road segment received at the gateway for a wide range of vehicle densities.

\section{Data Collection Challenges}\label{sec:chall}
The envisioned scenario is a city environment with moving cars which are collecting sensor data about the city, e.g. noise or air pollution, and the VANET is used as sensing infra-structure for gathering and sending the data to a back-office outside the VANET for monitoring purposes. This configures a many-to-one communication pattern, and it is assumed each node knows its own geographic position and that of the gateway.

In a city, cars move along the streets with unpredictable patterns and buildings block the communication between vehicles in non-aligned streets, creating a very volatile environment, where a single direct link lasts on average 20~s \cite{viriyasitavat_jsac2011}. Data collection protocols for wireless sensor networks and MANETs foresee some movement and provide mechanisms for path re-establishment, but they would need to be used too often in such volatile scenarios causing too high an overhead.

On the other hand, the amount of vehicles in each street segment is highly variable in time and across different streets, and not feasible to be estimated in real-time by vehicles in other streets with low overhead. This makes it difficult to use source routing to take the packet towards the destination, because it is not possible to know at the packet source, or anywhere along the way, whether there are vehicles that can serve as forwarders in the street segments between any vehicle and the gateway, causing packet die out with high likelihood. 

Furthermore, during traffic congestion, communication can be very difficult due to an overly loaded shared medium. In this situation, which is common in urban areas, requiring additional message exchange for coordination purposes increases further the medium load. 

A protocol for data collection in an urban scenario using vehicle-to-vehicle communication must take all these limitations into account. 

\section{Protocol Design}
\label{sec:data-gathering}
The proposed protocol is Urban Data Collector (UDC) protocol, a data gathering protocol that has the following features: 1) it does not require nodes to exchange periodic messages with their neighbors; 2) it uses the 802.11p MAC layer; 3) it takes advantage of redundant forwarding to increase packet delivery to the gateway; 4) it limits the amount of redundancy using suppression techniques.

The UDC protocol is a network layer protocol that can be run on top of the recently standardized 802.11p MAC. It is a broadcast- and receiver-based protocol, so that it can take advantage of any node that receives a forwarded packet without requiring exchange of information about node present in the neighborhood. This is especially relevant because the MAC layer does not provide acknowledgments, and it is not easy for a forwarder node to know whether its packet has been received by another node. 
Next we exemplify why some well-known solutions are not adequate for data gathering in urban areas, and explain how our protocol addresses those issues.

\begin{figure}[h]
\centering
\includegraphics[width=0.4\columnwidth]{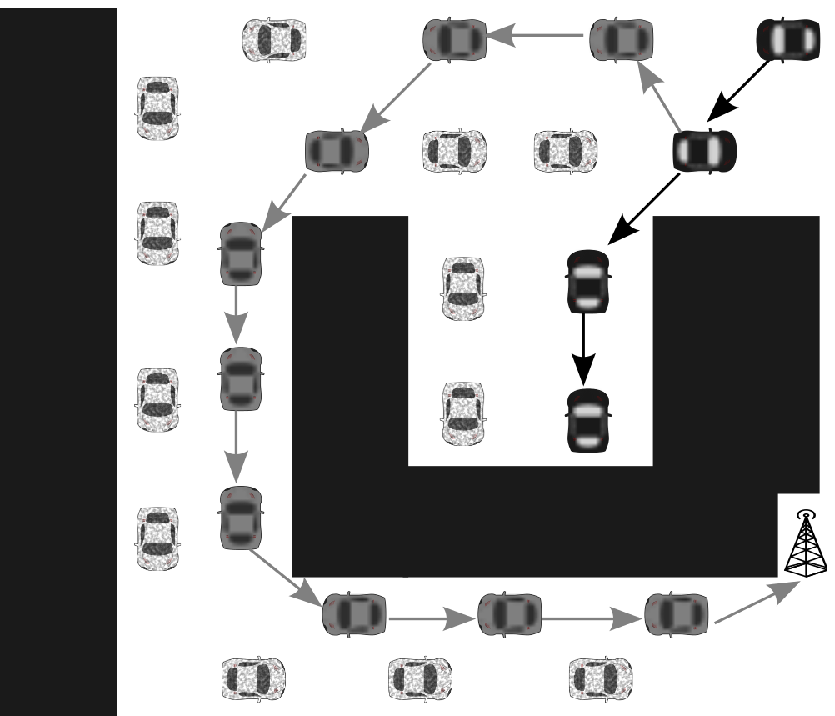}
\caption{UDC help packets to get to the gateway when a path in the direction of gateway is blocked by a building.}
\label{fig:ex1}
\end{figure}


Figure~\ref{fig:ex1} illustrates a case where a greedy protocol cannot deliver a packet to the gateway (shown as an antenna on the right bottom) because a chosen forwarder does not have any neighbor in the direction of the gateway. Protocols like GPSR~\cite{Karp2000} use perimeter routing technique to recover from the blocked path and find another path to the gateway, but these introduce a high delay and cause a great deal of additional transmissions. Because the proposed protocol follows two paths simultaneously (black and gray paths in Figure~\ref{fig:ex1}), although the black path is blocked, the gray path provides a redundant path which avoids packet die out.

\begin{figure}[h]
\centering
\includegraphics[width=0.4\columnwidth]{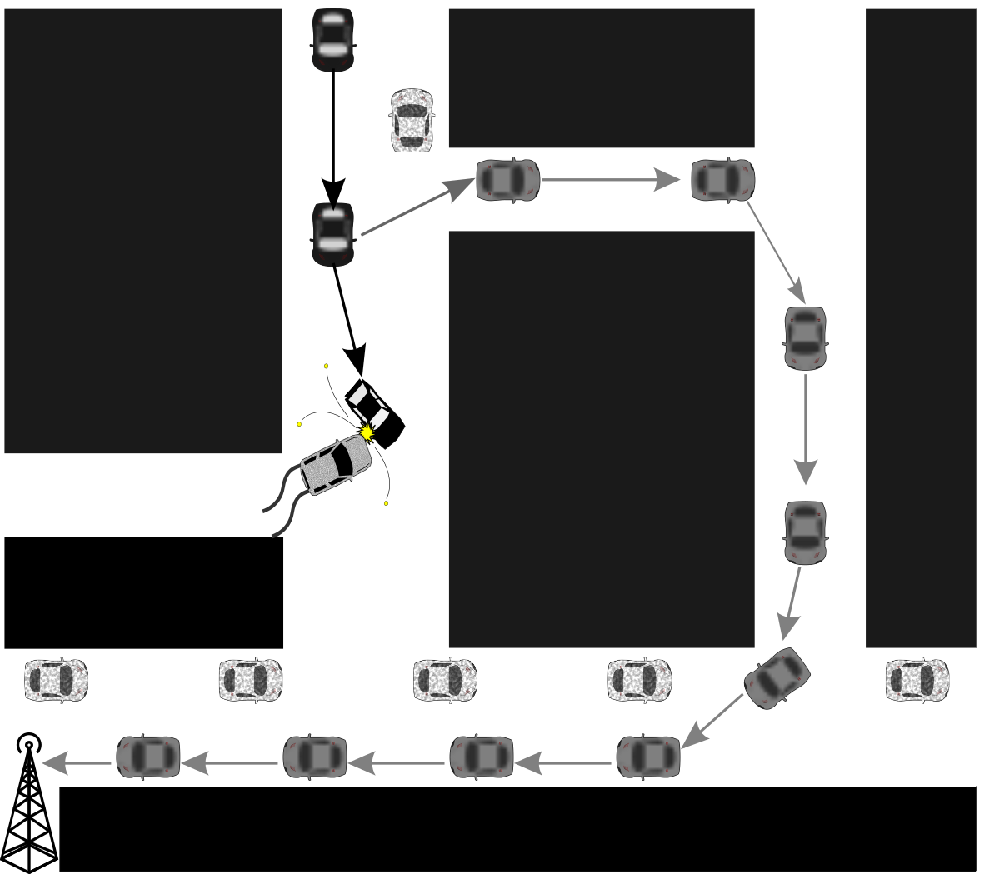}
\caption{UDC avoid packet die out when an accident happens on the direction of the gateway.}
\label{fig:ex2}
\end{figure}


In Figure~\ref{fig:ex2}, the main path (black path) to the gateway is blocked  because the packet reaches a  node that has no other nodes in range. This can happen because of the high volatility of the network, where the node that sent the packet is no longer in range after the routing dead-end has been detected. In this case, the perimeter technique will not be able to recover from this situation because there are no other nodes in the communication range of the single vehicle that has the packet. Because our protocol uses both the main path (black path) and the redundant path (gray path) for forwarding the packet the likelihood of both dying out is reduced.

Existing protocols get stuck in a blocked road or a road without forwarders toward the gateway because most of them take a binary decision for selecting the next forwarder. UDC gives the forwarding opportunity to several neighbor nodes to create redundant paths to avert packet die out. Finally, a fully distributed suppression technique that requires no coordination among nodes has been used to limit the amount of redundant paths so as to reduce the overhead. This is accomplished by taking advantage of two different types of forwarding: directional forwarding and probabilistic forwarding. Directional forwarding means that each node, upon receiving a packet, forwards it in the direction of the gateway after the expiration of a timer, which is used for giving forwarding priority to nodes that represent more progress towards the gateway. Probabilistic forwarding means that each node forwards the received packet with a probability which depends on difference distances between itself and the previous node to the gateway. The next sections describe in detail how these two techniques are combined to provide the desired functionality.

\subsection{Directional Forwarding}\label{sec:timer}
UDC uses a network layer timer calculated as a function of the difference between the distance of the previous and potential forwarder to the gateway to prioritize forwarders nearer to the gateway. The timer is calculated at each node in a distributed manner based only on information contained in the header of the received packet and of the node itself. 
\begin{figure}[h]
\centering
\includegraphics[width=\columnwidth]{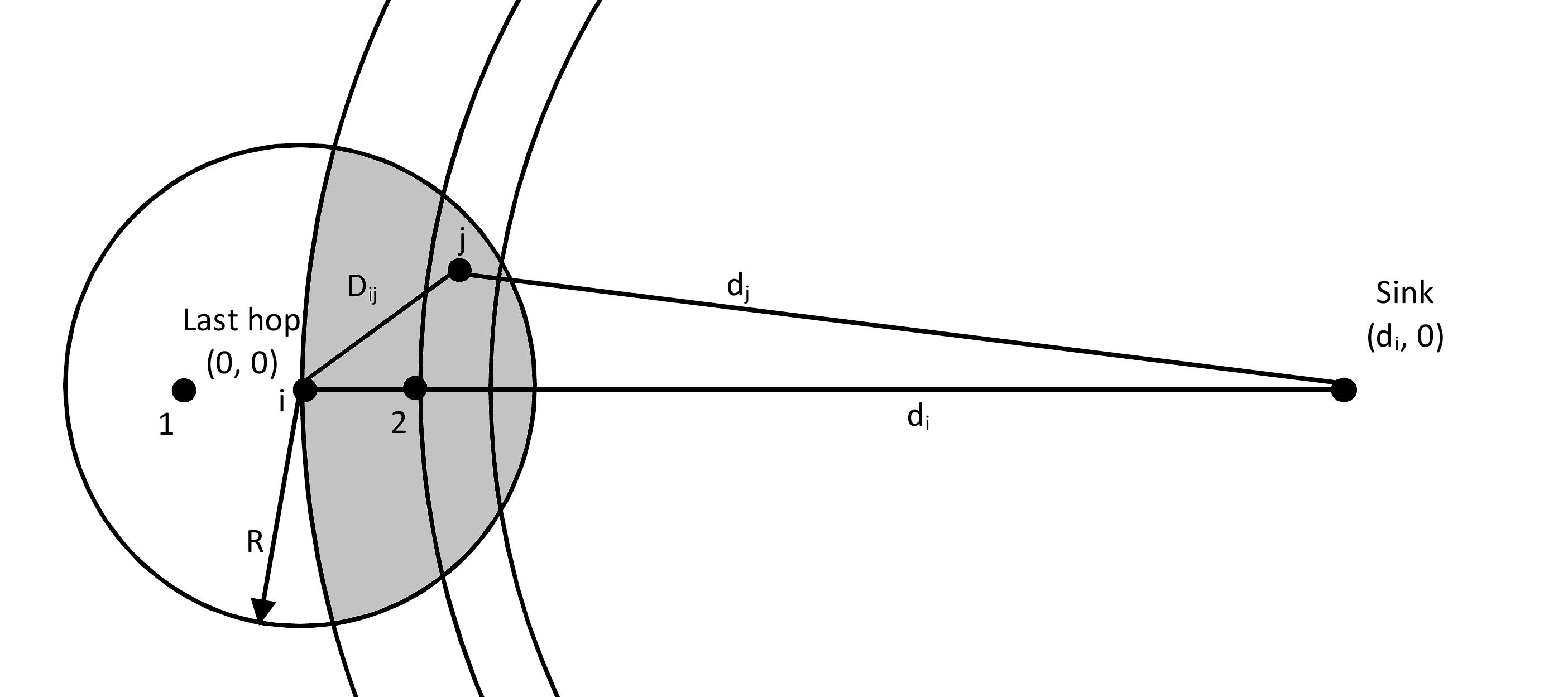}
\caption{One-hop forwarding entities}
\label{fig:entity}
\end{figure}

Figure~\ref{fig:entity} illustrates the scenario used to explain the calculation of the timer: node $i$ is the sender of the packet and nodes $1$, $2$ and $j$ are receivers and potential forwarders. The communication range is divided in two different areas, marked white and gray in Figure~\ref{fig:entity}. Nodes in the gray area are nearer to the gateway than the previous hop, whereas nodes in the white area are farther away. Nodes that represent the most progress towards the destination should be preferred forwarders. These are the nodes located closer to the end of the transmission range in the direction of the gateway, hence they shall have shorter timers and forward earlier than nodes farther away from the gateway. 

The time coefficient is calculated according to Equation~\ref{c2}
. This timer coefficient lies in range [0, 1] and is then multiplied by the maximum timer $T_{max}$ to deterministically calculate the actual timer value at all potential forwarders (Figure~\ref{fig:timer}).
\begin{equation}
\label{c2}
C = (1.0 + \frac{d_j - d_i - R}{2R}),
\end{equation}
where  $d_j$ is the distance of node $j$ to the gateway, $d_i$ the distance between the previous hop and the gateway, and $R$ is the communication range. 

\subsection{Suppression Techniques}
\label{sec:supp}
Suppression techniques are mechanisms that effect the reduction of the number of packets re-forwarded at each hop. The simplest suppression mode is that any node that hears a packet being forwarded while waiting for timer expiration stops the timer and discards the packet. In this way, the node thats correctly receives a packet and is closest to the gateway will be the first to forward, and any node that hears that forwarding will not double forward. But nodes on different sides of the previous hop may not hear each other, i.e., nodes in the white area do not hear nodes in the grey area in Figure~\ref{fig:entity} and cause duplicated forwarding. This problem could be solved by limiting the forwarder nodes to nodes that are nearer to the gateway than the previous hop. But this is likely to raise a problem for low node density, causing packet die out in the more likely event of lack of a forwarder nearer to the gateway.

Four different levels of suppression has been studied to overcome these problems: basic, weak, moderate and strong suppression. In the basic method, nodes discard a packet if they receive the same packet during the timer or wait time (see Section~\ref{sec:wait}), meaning that it has been forwarded by a node which is closer to the destination.
\begin{figure}[h]
\centering
\includegraphics[width=\columnwidth]{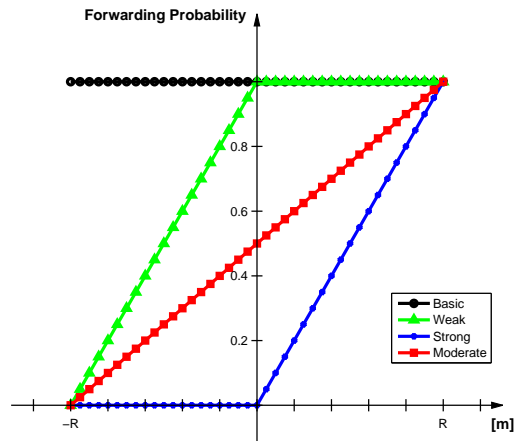}
\caption{Forwarding Probability calculated for different suppression techniques}
\label{prob-func}
\end{figure}

In weak suppression, UDC reduces the number of forwarders by using probabilistic forwarding for the nodes with higher distance to the gateway (white area in Figure~\ref{fig:conf}), while nodes closer to the destination (gray area in Figure~\ref{fig:conf}) forward the packet with probability one. Nodes in the white area forward a packet with a probability that decreases with increasing distance to the gateway, calculated according to Eq.~\ref{v2}. 
\begin{equation}
\label{v2}
p_{fwd} = (1.0 - \frac{d_j - d_i}{R})
\end{equation}

In strong suppression, nodes in the gray area, which are ranked by timer, forward the packet with different probability, calculated according to Eq.~\ref{v3} and nodes in the white area do not forward the packet at all.
\begin{equation}
\label{v3}
p_{fwd} = (\frac{d_i - d_j}{R})
\end{equation}

An intermediate level, moderate suppression, is added for which the forwarding probability is increased linearly with decreasing distance of potential forwarders to the gateway, calculated using Eq.~\ref{v4}.
\begin{equation}
\label{v4}
p_{fwd} = (\frac{d_i - d_j + R}{2R})
\end{equation}

Figure~\ref{prob-func} shows the forwarding probability of different suppression levels where the x-axis shows the difference distance between the last hop and potential forwarders to the gateway ($d_i - d_j$) relative to the communication range ($R$).

\begin{figure}[h]
\centering
\includegraphics[width=\columnwidth]{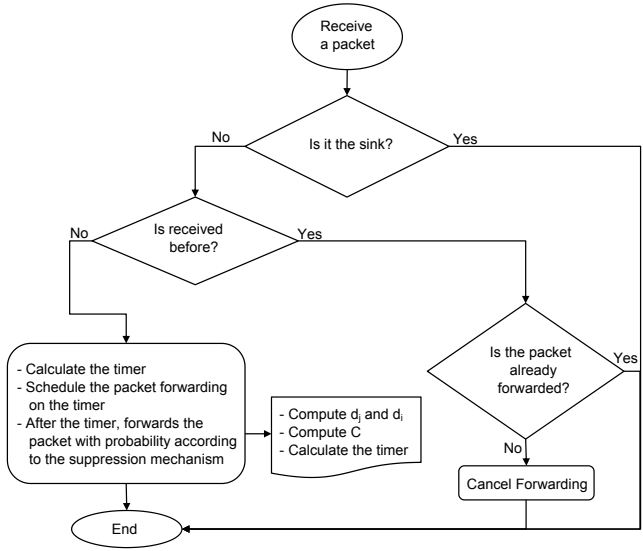}
\caption{Protocol algorithm}
\label{fig:algorithm}
\end{figure}

The complete network layer forwarding protocol algorithm is described in Figure~\ref{fig:algorithm}. Upon receiving a packet, node checks if it is not a gateway and consequently if it did not  receive the packet before, it calculates the timer (section~\ref{sec:timer}) and in the end of the timer according to the probability (section~\ref{sec:supp}) forwards the packet.

\subsection{Channel Access Time}\label{sec:wait}
\begin{figure}[h]
\centering
\includegraphics[width=\columnwidth]{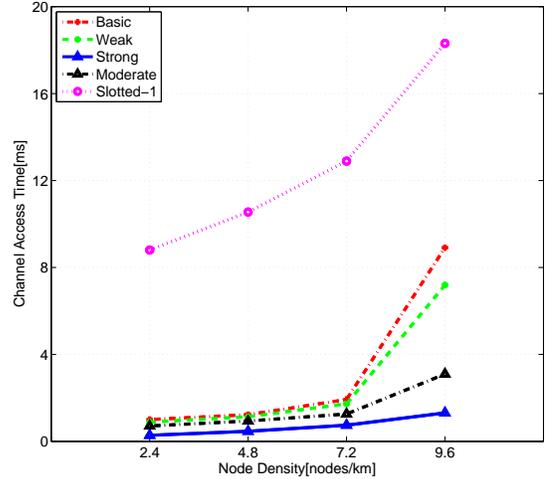}
\caption{Average channel access time for 4 suppression techniques.The slotted 1-persistence protocol used for comparison is introduced in Section~\ref{sec:rel-work}.
}
\label{fig:Allchtline}
\end{figure}

The average CSMA/CA channel access time obtained from simulation of UDC protocol with different suppression techniques is illustrated in Figure~\ref{fig:Allchtline}. The average channel access time for all node densities for 4 different suppression techniques was 5~ms
. This time is embedded in the protocol by letting each node wait for this duration before starting the timer at the network layer, to allow the suppression mechanisms to hear packets forwarded by other nodes.


\vspace{-0.05in}
\section{Simulation}
\label{sec:simulation}
The Network Simulator 3 (NS-3)~\cite{ns3} version 3.9 is used to simulate a $2.5\times2.5$~km Manhattan Grid topology with $5\times5$ roads, adding up to a total of $30 km$ road length. There is one gateway located in the center of the topology, at (1250~m, 1250~m), and all other nodes are data sources. The communication range is set to 500~m using a path-loss exponent of 2.5, and each node has on average at least four nodes and at most 84 nodes (crossroads with 42 nodes/km) within its communication range. The data rate is set to 1~kbps with100~Bytes packets, which is sufficient for envisioned sensing applications~\cite{Rodrigues2011}. The other simulation parameters can be found in Table~\ref{tbl:sim}.
\begin{table}[h]
\caption{Simulation Set up}
\centering
\begin{small}
\begin{tabular}{| c | c |}
\hline
\bf{Parameter Name} & \bf{Value}\\[1ex]
\hline\hline
Node Density & 4, 8, 16 and 42 nodes/km\\
Average Speed & 14~m/s\\
Minimum Speed & 3~m/s\\
\hline
MAC protocol & 802.11p\\
Channel Data rate & 6~Mbps\\
Channel Bandwidth & 10~MHz\\
Communication Range & 500~m\\
Propagation Model & Nakagami-m (m=1.56)\\
\hline
\end{tabular}
\end{small}
\label{tbl:sim}
\end{table}


Two seconds warm-up time is considered at the start of the simulation before reaching steady-state. The simulated time is set to guarantee that each node generates at least 20 packets. UDC with four different suppression levels (Section~\ref{sec:supp}) and the maximum timer equal to $T_{max}=5ms$ is used for simulation.
The slotted-1 persistence protocol~\cite{Wisitpongphan2007} is employed for comparison because it showed the best performance in a benchmark of existing protocols for data gathering using VANET in~\cite{NozariZarmehri2011} (see Section~\ref{sec:rel-work}).

\section{Performance Evaluation}
\label{sec:performance}
The performance of UDC is evaluated using the following metrics:
\begin{itemize}
\item \textbf{Sensing accuracy}: In the city environment for having an explicit view about the city, the urban sensing application needs having information about each road segment. Then the application can extract useful information about each road segment and expand it from road segment to entire city. So the sensing accuracy is defined to show how obtained information is accurate and can be applicable for the urban sensing. Therefore, it has been defined as number of received packets from each road segment in a second from all nodes within that section. To calculate it, after analyzing the collected data at the gateway, the received packets are separated depending on the source`s location.
\item \textbf{Network efficiency}: The question of how well the protocol acts in the broadcast fashion is defined as network efficiency. This metric shows how many of the packet forwardings are used for the packets which get to the sink. The defination is: the number of hops for each packet receives at the sink, including both unique packets and their replicas, divided by the number of all forwarded packets in all nodes.
\end{itemize}
Additionally, extensive simulations were run to determine the different reasons for the packet discards at each node, and the evaluation of these results is shown on this section.

\subsection{Sensing accuracy}
\label{sec:sensing}
\begin{figure}[h]
\centering
\includegraphics[width=\columnwidth]{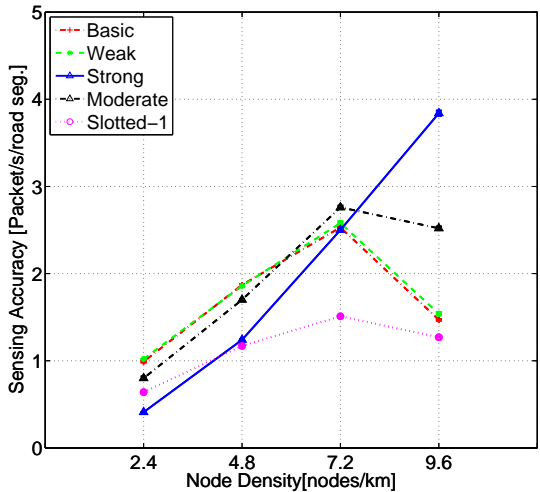} 
\caption{Sensing accuracy for different data collection mechanism in urban area}
\label{fig:SA}
\end{figure}

Figure~\ref{fig:SA} shows the sensing accuracy for the UDC with different suppression levels. It plots the average number of received packets from each road segment at the gateway in one second windows in the simulated road topology against varying node densities. It shows that the sensing accuracy is most dependent on the node density when using the strong suppression level. For the lowest node density, it discards more than 87\% of the received packets at each hop and is not able to produce enough redundancy paths, causing high probability of packet die out. As the node density increases, the strong suppression level represses about 97\% of received packets at each hop and thus reduces the number of collisions, achieving the highest sensing accuracy for the highest node density. 

On the other hand, the weak and basic suppression level achieve similar sensing accuracy with less dependency to the node density. Since the sensing application needs to obtain information from each road segment independently of the node density, a protocol with less variations in the sensing accuracy for different node densities is a better choice for a sensing application. This fact and the fact that both provide the highest accuracy for the lowest node density, makes them a better choice for the envisioned application. The moderate suppression level has a sensing accuracy that lies midway between the performance of weak and strong, showing better performance than strong suppression at low node density and better performance than weak suppression at high node density.

\subsection{Network Efficiency}
\begin{figure}[h]
\begin{center}
\leavevmode
\includegraphics[width=\columnwidth]{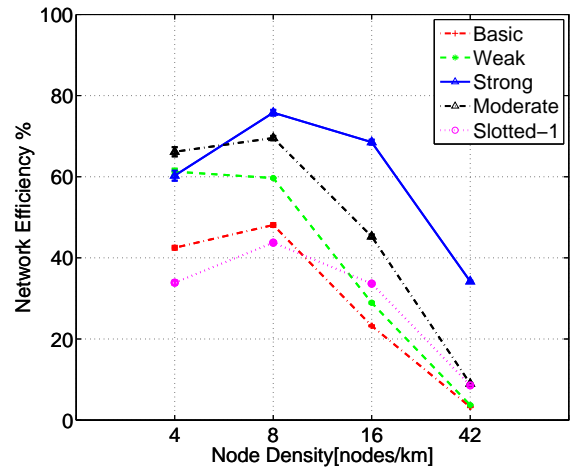} 
\caption{Efficiency for using different suppression techniques}
\label{Efficiency}
\end{center}
\end{figure}

Figure~\ref{Efficiency} shows that the UDC with strong suppression level uses broadcast forwarding in the most efficient way, since a larger percentage of the packets forwarded in the network end up reaching the sink. More interestingly, the weak suppression level shows a higher network efficiency than the basic level while achieving similar sensing accuracy (see previous section) since the lower amount of forwarded packets causes a lower number of collisions. The moderate suppression level again lies midway between strong and weak suppression levels, as intended at its construction.
These results also show that the efficiency of the protocol drops with increasing node density due to the high amount of collision caused by a higher medium load, as expected for a broadcast protocol in a shared medium.

\vspace{-0.05in}

\subsection{Packet Drop Analysis}\label{sec:drop}
\begin{figure}[h]
\centering
\includegraphics[width=\columnwidth]{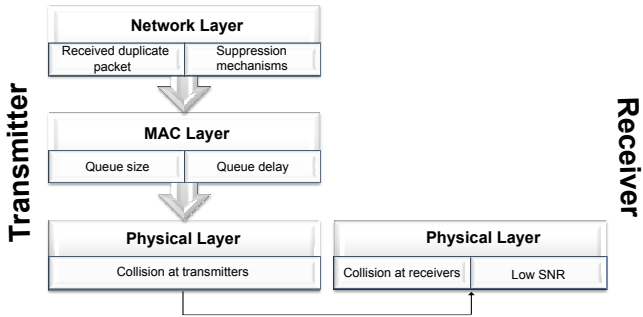}
\caption{Different reasons for drop packets depend on different communication layers}
\label{reasons}
\end{figure}

An extensive simulation is carried out to analyze the reasons of performance drop for increasing node density, with the aim of gaining insight into the possible ways to improve the performance of the protocol.
Figure~\ref{reasons} identifies the different possible reasons for packet discards at the different communication layers.


In this analysis, only the packets that do not get to the sink is considered and the drop reasons for these packets is counted. Table~\ref{tbl:drop} shows
 the reason why the last replication of a packet still alive in the network dies out before reaching the gateway. This table also shows that the main reason for dropping the packets is excessive suppression, i.e. suppression of packets that would have been useful for reaching a higher sensing accuracy. 

Also, even for low node densities, suppression and collision are the main causes of packet die out, and not low SNR, i.~e. packets do not die out because of lack of forwarders, but due to the broadcast nature of the protocol, which causes collisions and requires suppression. On the other hand, the single main cause for packet die out for the strong method is suppression, for any node density. 

Moreover, we conclude that an intermediate solution between the weak and strong suppression techniques would achieve a better balance between the collision and suppression trade-offs. So UDC with moderate suppression level is proposed as the best compromise, in accordance with the results of the performance evaluation in Section~\ref{sec:performance}.

\begin{table}[!ht]
\caption{Percentage of each reasons in the protocol performance for different node densities}
\centering
\begin{tabular}{| c | c | c | c | c | c |}
\hline
\bf{Density} & \bf{Sup.} & \bf{Sup.} & \bf{Col.} & \bf{Low} & \bf{PDR}\\[1ex]
\bf{nod./km} & \bf{Method} & \bf{\%} & \bf{\%} & \bf{SNR\%} & \bf{\%}\\
\hline\hline
4 & Strong  & 81.06 & 0 & 0.46 & 18.48 \\
& Weak & 19.23 & 6.15 & 1.4 & 73.22 \\
\hline
8 & Strong & 48.01 & 0.02 & 0.04 & 51.93 \\
& Weak & 15.73 & 6.92 & 0.21 & 77.14 \\
\hline
16 & Strong & 39.11 & 0.22 & 0.02 & 60.65 \\
& Weak & 28.64 & 11.93 & 0 & 59.43 \\
\hline
42 & Strong & 55.27 & 0.79 & 0 & 43.94 \\
& Weak & 53.79 & 33.59 & 0 & 12.62 \\
\hline
\end{tabular}
\label{tbl:drop}
\end{table}

\vspace{-0.1in}
\section{Related Work}
\label{sec:rel-work}
Existing solutions for VANET sensing either apply on-demand querying for local dissemination
~\cite{lee_IEEETVT2009}, sometimes keeping the data in the location
~\cite{dikaiakos_JSACA2007}, or rely on delay-tolerant networking and open Wi-Fi access points for sending the data to the Internet backbone~\cite{Hull_SenSys2006}. The first are inefficient for real-time
 monitoring due to the query overhead and the need to globally access the data, and the latter cannot guarantee up-to-date data. 

On the network level, a wide range of broadcast-based vehicle-to-vehicle unicast routing protocols has been proposed and we categorize them in two classes: sender-oriented and receiver-oriented. Sender-oriented protocols locally exchange beacons to obtain information. It enables the current forwarder to select the next forwarder node among its neighbors. This is the approach taken by protocols like Greedy Perimeter Stateless Routing (GPSR)~\cite{Karp2000}, Position-based Multi-hop Broadcast Protocol(PMBP)~\cite{Bi2009}, Emergency Message Dissemination for Vehicular environments (EMDV) ~\cite{Torrent-Moreno2009}, or Cross Layer Broadcast Protocol(CLBP)~\cite{Bi2010}. Sender-oriented protocols are inefficient for data gathering in urban areas due to the high overhead of continuously exchanging beacons and because additional mechanisms must be added to verify at each hop whether the chosen forwarder actually received and forwarded the packet. Furthermore, explicitly choosing a single forwarder in urban scenarios is not adequate for urban data collection, as we discussed in Section~\ref{sec:chall}.

Receiver-oriented protocols forward data packets without prior message exchange and employ mechanisms to reduce the network overhead. Protocols like Contention-based forwarding (CBF)~\cite{Fusler2003} or Efficient Directional Broadcast (EDB) ~\cite{Li2007} use contention-based forwarding without handshaking. Upon receiving a packet, all nodes wait for a timer proportional to the geographic progress towards the gateway, but both require the exchange of further messages per packet and hop. 
CBF uses request-to-forward (RTF) and clear-to-forward (CTF) messages to reduce duplicate packets and EDB sends an ACK message before forwarding. 

Finally, 
\cite{Wisitpongphan2007} proposed 
three broadcast-based protocols that use basic per-hop forwarding and suppression techniques for data dissemination: Weighted p-Persistence, Slotted-1 Persistence and Slotted p-Persistence broadcasting. We evaluated the performance of these protocols for data collection over VANET~\cite{NozariZarmehri2011} and the Slotted-1 Persistence protocol shows better performance than the two others. Hence, we used it for comparison and introduce it here.

Slotted-1 persistence uses probabilistic and slotted forwarding. Each node $j$, upon receiving a packet from node $i$, re-broadcasts the packet at a timeslot $T_{S_{ij}}$ if it receives the packets for the first time and does not receive any duplicates before the assigned timeslot, otherwise it discards the packets. $T_{S_{ij}}$ is calculated by $T_{S_{ij}} = S_{ij} \times{} \tau$, 
where $\tau$ is the estimated one hop delay and \textit{$S_{ij}$} is the assigned slot number, calculated by:
\begin{equation}
\label{pij}
  S_{ij} = \left\{ 
  \begin{array}{l l}
   N_s \times{} (1-\frac{D_{ij}}{R}) & \quad {D_{ij} \leq R} \\
  \\
   0 & \quad {D_{ij} > R}\\
  \end{array}, \right.
\end{equation}
where $N_s$ is the predetermined number of slots, $D_{ij}$ is the distance between nodes $i$ and $j$ and $R$ is average communication range.

\section{Conclusion and Future Work}
\label{sec:conclusions}
We envision the usage of a VANET as infra-structure for an urban cyber-physical system that makes up-to-date data about various parameters of an urban area available to services outside of the network. We propose and evaluate the use of UDC protocol, a broadcast- and receiver-based forwarding protocol.

The effect of four different suppression techniques on the sensing accuracy and network overhead is evaluated using NS-3 large scale simulation. The results reveal that for supporting the sensing applications in the urban area, the weak suppression increases the sensing accuracy with lower dependency to the node density. Also it shows that the weak method has less excessive suppression than other methods and is the best solution of those evaluated for urban sensing.

As part of ongoing efforts, the study of network limitation and its effect on the sensing accuracy will be investigated. Subsequently, the effect of the number of gateway on the sensing accuracy will be explored.

\bibliographystyle{abbrv}
\small{
    \bibliography{ref}
}

\end{document}